\documentclass[aip,jmp,amsmath,amssymb,reprint]{revtex4-1}
\usepackage{graphicx}
\usepackage{dcolumn}
\usepackage{bm}
\usepackage[utf8]{inputenc}
\usepackage[T1]{fontenc}
\usepackage{mathptmx}
\begin{document}

\title{Particle Production Between Isometric Frames on a Poincaré Patch of $\text{AdS}_2$}

\author{Joao Paulo M. Pitelli}
 \altaffiliation[Also at ]{The Enrico Fermi Institute, The University of Chicago,
Chicago, IL.}
 \email{pitelli@ime.unicamp.br}
\affiliation{ 
Departamento de Matemática Aplicada, Universidade Estadual de Campinas,
13083-859 Campinas, São Paulo, Brazil
}%

\author{Vitor S. Barroso}
 \email{barrosov@ifi.unicamp.br}
\affiliation{Instituto de F\'isica ``Gleb Wataghin'', Universidade Estadual de Campinas 13083-859 Campinas, São Paulo, Brasil}%

\date{\today}

\begin{abstract}
In a recent paper [J. P. M. Pitelli, Phys. Rev. D {\bf 99}, 108701 (2019)], one of us showed that the vacuum state associated to conformal fields on a Poicaré patch of anti-de Sitter spacetime is not $\text{AdS}$ invariant for fields satisfying non-trivial boundary conditions (by non-trivial we mean neither Dirichlet nor Neumann) at the conformal boundary. In this way, two isometrically related observers in anti-de Sitter space have different notions of no particle content. Therefore, an observer who is suddenly transported to a different (but isometrically related) frame will feel a bath of particles. This process contradicts our intuitive notion based on our experience in Minkowski spacetime, where the vacuum is Lorentz invariant, and no particle is produced between boosted frames. We show that the total number of produced particles is finite, but grows without limit when we approach (via isometric transformation) Dirichlet or Neumann boundary conditions since in these cases the vacuum is invariant. 

\end{abstract}

\maketitle

\section*{\label{sec:intro}Introduction}

It is well known that two inertial observers in Minkowski spacetime have the same notion of no particle content. The vacuum state in these frames is Lorentz invariant and uniquely defined given the global hyperbolicity of the spacetime.  In this way, to experience non-trivial effects in the absence of sources/matter, we need an accelerated observer. This (non-inertial) observer feels a thermal bath of particles with a temperature proportional to the acceleration (Unruh effect~\cite{unruh1}).  This situation changes dramatically in non-globally hyperbolic spacetimes. Since there is no Cauchy surface, the evolution of classical fields may not be uniquely defined given the initial data. In addition to the timelike Killing field, which determines the positive energy solutions, one should specify a boundary condition at the edge of spacetime. This edge can be a naked singularity, like the one found on a conical spacetime~\cite{pitelli1, konkowski1}, or even a timelike boundary as the one found in $\textrm{AdS}$~\cite{pitelli2,pitelli3,pitelli4}. In this way, for each boundary condition, there is an associated vacuum state. One crucial question is whether this vacuum state satisfies the symmetries of the background spacetime. If this is not the case, then observers who are isometrically related may have different notions of particle content.

In Refs.~\onlinecite{pitelli2,pitelli3}, the authors showed that the vacuum might not be invariant under isometric transformations, even in a maximally symmetric spacetime, like anti-de Sitter space. For a conformal field on a Poincar\'e patch of $\textrm{AdS}_2$ (denoted $\textrm{PAdS}_2$), this break of invariance can be explained in the following way\cite{dappiaggi}: given the metric
\begin{equation}
ds^2=\frac{-dt^2+dz^2}{z^2},\,\,\,z>0
\label{metric}
\end{equation}
on the patch,  the Klein-Gordon equation reduces to the wave equation
\begin{equation}
\frac{\partial^2 \varphi(t,z)}{\partial t^2}=\frac{\partial^2 \varphi(t,z)}{\partial z^2},\,\,\,z>0. 
\label{wave}
\end{equation}
To solve this equation, one also needs a boundary condition. One suitable choice is the Robin boundary conditions, i.e.,
 \begin{equation}
\varphi(t,z=0)-\beta \frac{\partial\varphi(t,z=0)}{\partial z}=0,
\label{bc}
\end{equation}
since they provide the self-adjoint extensions of the spatial component of the wave equation, which ensures a sensible dynamical evolution for the field~\cite{wald,wald2}. In Eq.~(\ref{bc}), $\beta$ is a parameter with the dimension of length. At this point, it becomes clear that the Klein-Gordon equation in $\textrm{PAdS}_2$ is equivalent to the wave equation in $\mathbb{R}\times(0,\infty)$ endowed with the Minkowski metric. Since there is only one global symmetry in this case, namely time invariance, there is no guarantee that the other symmetries in $\textrm{PAdS}_2$ will be respected. Nevertheless, there are two cases in which $\textrm{AdS}$
invariance is preserved, namely when $\beta=0$ (Dirichlet) and $\beta=\infty$ (Neumman). In these cases, the boundary conditions do not introduce any additional length scale. For $0<\beta<\infty$, the vacuum is not $\textrm{AdS}$ invariant, and we expect that the notion of particle content will change even when performing a symmetry transformation.

 We consider a  dynamical process in which an observer  initially in a frame $(t,z)$ suddenly (at $t=t^\prime=0$)  switches to another frame $(t^\prime, z^{\prime})$ related to the first one by an isometric transformation. We will see that, in contrast to what happens in Minkowski spacetime, where two inertial frames share the same vacuum state and a boosted observer does not feel any quantum field effect, the observer in $\textrm{PAdS}_2$ will feel a bath of particles.  As will be displayed later, the role of this instantaneous symmetry transformation on the field $\varphi$ is that of changing the boundary condition at the conformal boundary. Since each boundary condition gives a particular notion of vacuum, this sudden change of boundary condition will be responsible for the creation of particles. A similar situation occurs when there is an instantaneous change in the boundary condition satisfied by a scalar field in $(1+1)$-dimensional Minkowski spacetime with finite spatial interval. This setup mimics the emergence of a singularity at each spatial boundary (chosen as $x=0$ and $x=\pi$  w.l.o.g.). In Ref.~\onlinecite{ishibashi}, Ishibashi and Hosoya studied this scenario,  showing  that if at $t=0$ a scalar field $\varphi$ changes from Neumann (N) to Dirichlet (D) boundary condition at the spatial boundaries, then an infinite number of particles is produced so that the in and out vacua are not unitarily equivalent. In Ref.~\onlinecite{miyamoto}, the same effects were investigated in $\mathbb{R}\times[0,\infty)$, i.e., Minkowski spacetime with a semi-infinite spatial component. Miyamoto showed that both the $\text{N}\to\text{D}$ and $\text{D}\to\text{N}$  cases also lead to an infinite number of produced particles within this setup.

In the two references above-mentioned, the authors argue that the sudden change of boundary condition may mimic the creation of a naked singularity. Here, we investigate the production of particles due to a change in boundary conditions in a completely different (and, probably, more physical) situation. We show that  an isometric transformation in $\textrm{PAdS}_2$ at $t=0$ changes the (Robin) boundary condition respected by the scalar from $\beta$ to $\beta^{\prime}$ (where $\beta^{\prime}$ is not arbitrary but is related to $\beta$ by the isometry). We compute precisely the total number of particles produced in this sudden change of frames, revealing that it is finite when $0<\beta,\beta^{\prime}<\infty$ (when the vacuum is not $\textrm{AdS}$ invariant) and tends to infinite in the limits $\beta^{\prime}\to 0$ or $\beta^{\prime}\to\infty$. The latter follows from the change between a non-$\textrm{AdS}$ invariant vacuum to an $\textrm{AdS}$ invariant one.

\section*{Change of Frames}

There are three Killing fields on $\textrm{PAdS}_2$, namely
\begin{equation}
\begin{aligned}
&\xi_1=\partial_t,\\
&\xi_2=t\partial_t+z\partial_z,\\
&\xi_3=(t^2+z^2)\partial_t+2t z\partial_z.
\end{aligned}
\label{killing}
\end{equation}
The flows of these fields generate coordinate transformations preserving the metric~(\ref{metric}) as well as the conformal boundary. Clearly, the wave equation~(\ref{wave}) is invariant under these transformations. However, to solve this equation we need a boundary condition of the form~(\ref{bc}), which fixes a preferred frame. Since $\xi_2$ and $\xi_3$ in~(\ref{killing})  transform the spatial coordinate, symmetries transformations generated by them will change the boundary condition at the conformal boundary. It is straightforward to see that the coordinate transformation due to $\xi_3$ would drastically change the boundary condition from the Robin form to a time dependent one. Conversely, the coordinate transformation generated by $\xi_2$ is simpler and given by
\begin{equation}
\begin{aligned}
t^{\prime}=\lambda t,\\
z^{\prime}=\lambda z,
\end{aligned}
\label{frames}
\end{equation}
with $\lambda>0$. This change of coordinates modifies the boundary condition~(\ref{bc}) to
\begin{equation}
\varphi(t^{\prime},z^{\prime}=0)-\beta\lambda\frac{\partial \varphi(t^{\prime},z^{\prime}=0)}{\partial z^{\prime}}=0.
\label{bc2}
\end{equation}
In this way, if $|0\rangle_\beta$ is the vacuum for the frame $(t,z)$, then the natural vacuum for the transformed frame $(t^{\prime},z^{\prime})$ is given by $|0\rangle_{\beta\lambda}$. These two vacua are different and we will see how they are related by studying the Bogoliubov coefficients between both representations $\varphi^{\beta}$ and $\varphi^{\beta\lambda}$ for the field respecting the wave equation~(\ref{wave}).

\section*{Bogoliubov coefficients}

In this section, we will calculate the Bogoliubov coefficients which appear due to a sudden (at $t=t^\prime=0$) change of frames $(t,z)\to (t^{\prime}, z^{\prime})$. This setting is equivalent to an instantaneous change of boundary condition satisfied by the scalar field at the conformal boundary $z=z^{\prime}=0$. We will see that this sudden change of boundary condition mixes the positive and negative frequencies solutions (with respect to the same Killing field $\xi_1$), being responsible to the production of particles at $t=0$.

Given the wave equation~(\ref{wave}) and a boundary condition of the form~(\ref{bc}) for the conformal scalar field  in $\textrm{PAdS}_2$ for $t<0$, it can be readily checked that $\varphi$ can be expanded in the form
\begin{equation}
\varphi(t,z)=\int_{0}^{\infty}{\left(a_{\omega}u_{\omega}^{(\beta)}(t,z)+a_{\omega}^{\dagger}u_{\omega}^{(\beta)\ast}(t,z)\right)d\omega},
\end{equation}
with the complete set of modes $\{u_{\omega}^{(\beta)}(t,z)\}$  
\begin{equation}
u_{\omega}^{(\beta)}(t,z)=\frac{1}{\sqrt{\pi\omega}}\frac{\sin{\omega z+\beta\omega\cos{\omega z}}}{\sqrt{1+\beta^2\omega^2}}e^{-i\omega t}
\end{equation}
being orthonormal in the Klein-Gordon inner product
\begin{equation}
\left(f,g\right)=-i\int_{0}^{\infty}{f(t,z)\overleftrightarrow{\partial_t}g^{\ast}(t,z)dz}.
\end{equation}

After the change of frames $\Lambda:x=(t,z)\to x^{\prime}=(t^{\prime},z^{\prime})$   at $t=0$ given by Eq.~(\ref{frames}), the modes transform as
\begin{equation}
u^{(\beta)}_{\omega}(x)\to u^{(\beta)^\prime}_\omega(x)= u^{(\beta)}_\omega(\Lambda^{-1}x).
\end{equation}
The field $\varphi$ become (for $t>0$)
\begin{equation}
\begin{aligned}
&\varphi^\prime(t,z)=\int_{0}^{\infty}{\left[a_{\omega}u_{\omega}^{(\beta)}\left(\frac{t}{\lambda},\frac{z}{\lambda}\right)+a_{\omega}^{\dagger}u_{\omega}^{(\beta)\ast}\left(\frac{t}{\lambda},\frac{z}{\lambda}\right)\right]d\omega}\\&
=\int_0^{\infty}{a_{\omega}\frac{1}{\sqrt{\pi\omega}}\frac{\sin{\omega \frac{z}{\lambda}+\beta\omega\cos{\omega \frac{z}{\lambda}}}}{\sqrt{1+\beta^2\omega^2}}e^{-i\frac{\omega}{\lambda} t}d\omega}+\text{c.c.},
\end{aligned}
\end{equation}
where c.c. means complex conjugate. By redefining $\omega\to\lambda\tilde{\omega}$ in the above equation, it becomes clear that this equation is equivalent to  a change in boundary conditions $\beta\to\beta\lambda$, i.e., 
\begin{equation}\begin{aligned}
\varphi^\prime(t,z)&=\int_0^{\infty}{\frac{\sqrt{\lambda}a_{\lambda\tilde{\omega}}}{\sqrt{\pi\tilde{\omega}}}\frac{\sin{\tilde{\omega} z+\beta\lambda\tilde{\omega}\cos{\tilde{\omega} z}}}{\sqrt{1+\beta^2\lambda^2\tilde{\omega}^2}}e^{-i\tilde{\omega}  t}d\tilde{\omega}}+\text{c.c.}\\&\equiv
\int_{0}^{\infty}{\left[b_{\tilde{\omega}}u_{\tilde{\omega}}^{(\beta\lambda)}(t,z)+b_{\tilde{\omega}}^{\dagger}u_{\tilde{\omega}}^{(\beta\lambda)\ast}(t,z)\right]d\tilde{\omega}}.
\end{aligned}\end{equation}

Following Ref.~\onlinecite{wald}, given an extension $A_{\tilde{\beta}}$ (corresponding to a boundary condition represented by the parameter $\tilde{\beta}$ in~(\ref{bc})) of the spatial part of the wave operator, namely $A=-\partial^2/\partial z^2$, the solution of the wave equation can be found through the relation
\begin{equation}
\varphi^{\tilde{\beta}}(t)=\cos{\left(\sqrt{A_{\tilde{\beta}}}t\right)}\varphi(0)+\frac{1}{\sqrt{A_{\tilde{\beta}}}}\sin{\left(\sqrt{A_{\tilde{\beta}}}t\right)}\frac{\partial}{\partial t}\varphi(0),
\label{expression for the field}
\end{equation}
with $\varphi(0)$ and $\frac{\partial}{\partial t}\varphi(0)$ being the initial data.

Our dynamical process can be described in the following way: Suppose  that at $t=0$, an observer initially at the $(t,z)$ frame is transported to the $(t^{\prime}, z^{\prime})$ frame given by~(\ref{frames}). Before $t=0$, positive frequency solutions satisfied~(\ref{bc}). After $t=0$, they satisfy~(\ref{bc2}). It is straightforward to check that
\begin{equation}
u_{\omega}^{(\beta)}(0,z)=\int_{0}^{\infty}{g(\omega,\tilde{\omega})u_{\tilde{\omega}}^{(\beta\lambda)}(0,z)d\tilde{\omega}},
\end{equation}
with
\begin{widetext}
\begin{equation}\begin{aligned}
g(\omega,\tilde{\omega})&=2\tilde{\omega}\int_{0}^{\infty}{u_{\omega}^{\beta}(0,z)u_{\tilde{\omega}}^{\beta\lambda\ast}(0,z)dz}\\&=\int_{0}^{\infty}{\frac{1}{\sqrt{\pi\omega}}\frac{\sin{\omega z+\beta\omega\cos{\omega z}}}{\sqrt{1+\beta^2\omega^2}}\frac{1}{\sqrt{\pi\tilde{\omega}}}\frac{\sin{\tilde{\omega} z+\beta\lambda\tilde{\omega}\cos{\tilde{\omega} z}}}{\sqrt{1+\beta^2\lambda^2\tilde{\omega}^2}}dz}\\&=
\frac{1+\lambda \beta^2\omega^2}{\sqrt{1+\beta^2\omega^2}\sqrt{1+\beta^2\lambda^2\omega^2}}\delta(\omega-\tilde{\omega})-\frac{\beta}{\pi}\frac{1}{\sqrt{1+\beta^2\omega^2}\sqrt{1+\beta^2\lambda^2\tilde{\omega}^2}}\sqrt{\frac{\tilde{\omega}}{\omega}}(1-\lambda)\frac{\omega\tilde{\omega}}{\omega^2-\tilde{\omega}^2}.
\end{aligned}
\end{equation}
\end{widetext}
The modes $u_{\omega}^{\lambda}$  at late times are then given by~(\ref{expression for the field}) with $\tilde{\beta}=\beta\lambda$ and initial conditions
\begin{equation}
u_{\omega}^{(\text{in})}(0,z)=u_{\omega}^{(\beta)}(0,z)=\int_{0}^{\infty}{g(\omega,\tilde{\omega})u_{\tilde{\omega}}^{(\beta\lambda)}(0,z)d\tilde{\omega}},
\label{initial1}
\end{equation}
and
\begin{equation}
\frac{\partial}{\partial t}u_{\omega}^{(\text{in})}(0,z)=-i \omega u_{\omega}^{(\text{in})}(0,z).
\label{initial2}
\end{equation}
By plugging Eqs.~(\ref{initial1}) and~\eqref{initial2} in Eq.~(\ref{expression for the field}), we find that
\begin{widetext}
\begin{equation}\begin{aligned}
u_{\omega}^{(\beta)}(t,z)&=\int_{0}^{\infty}{g(\omega,\tilde{\omega})\left[\cos{(\tilde{\omega}t)}u_{\tilde{\omega}}^{(\lambda\beta)}(0,z)-i\frac{\omega}{\tilde{\omega}}\sin{(\tilde{\omega}t)}u_{\tilde{\omega}}^{(\lambda\beta)\ast}(0,z)\right]d\tilde{\omega}}\\&=\int_{0}^{\infty}{\frac{g(\omega,\tilde{\omega})}{2}u_{\tilde{\omega}}^{(\lambda\beta)}(0,z)\left[\left(1+\frac{\omega}{\tilde{\omega}}\right)e^{-i\tilde{\omega} t}+\left(1-\frac{\omega}{\tilde{\omega}}\right)e^{i\tilde{\omega} t}\right]d\tilde{\omega}}\\&\equiv \int_{0}^{\infty}{\left(\alpha_{\omega\tilde{\omega}}u^{(\beta\lambda)}_{\tilde{\omega}}(t,z)+\gamma_{\omega\tilde{\omega}}u^{(\beta\lambda)\ast}_{\tilde{\omega}}(t,z)\right)d\tilde{\omega}},
\end{aligned}
\end{equation}
where $\alpha_{\omega\tilde{\omega}}$ and $\gamma_{\omega\tilde{\omega}}$ are the Bogoliubov coefficients given by
\begin{equation}
\begin{aligned}
&\alpha_{\omega\tilde{\omega}}=\frac{1+\lambda \beta^2\omega^2}{\sqrt{1+\beta^2\omega^2}\sqrt{1+\beta^2\lambda^2\omega^2}}\delta(\omega-\tilde{\omega})+\frac{2\beta}{\pi}\frac{1}{\sqrt{1+\beta^2\omega^2}\sqrt{1+\beta^2\lambda^2\tilde{\omega}^2}}(1-\lambda)\frac{\sqrt{\tilde{\omega}\omega}}{\omega-\tilde{\omega}},\\
&\gamma_{\omega\tilde{\omega}}=\frac{2\beta}{\pi}\frac{1}{\sqrt{1+\beta^2\omega^2}\sqrt{1+\beta^2\lambda^2\tilde{\omega}^2}}(1-\lambda)\frac{\sqrt{\tilde{\omega}\omega}}{\omega+\tilde{\omega}}.
\end{aligned}
\end{equation}
\end{widetext}

\begin{figure}[h]
\includegraphics[scale=0.4]{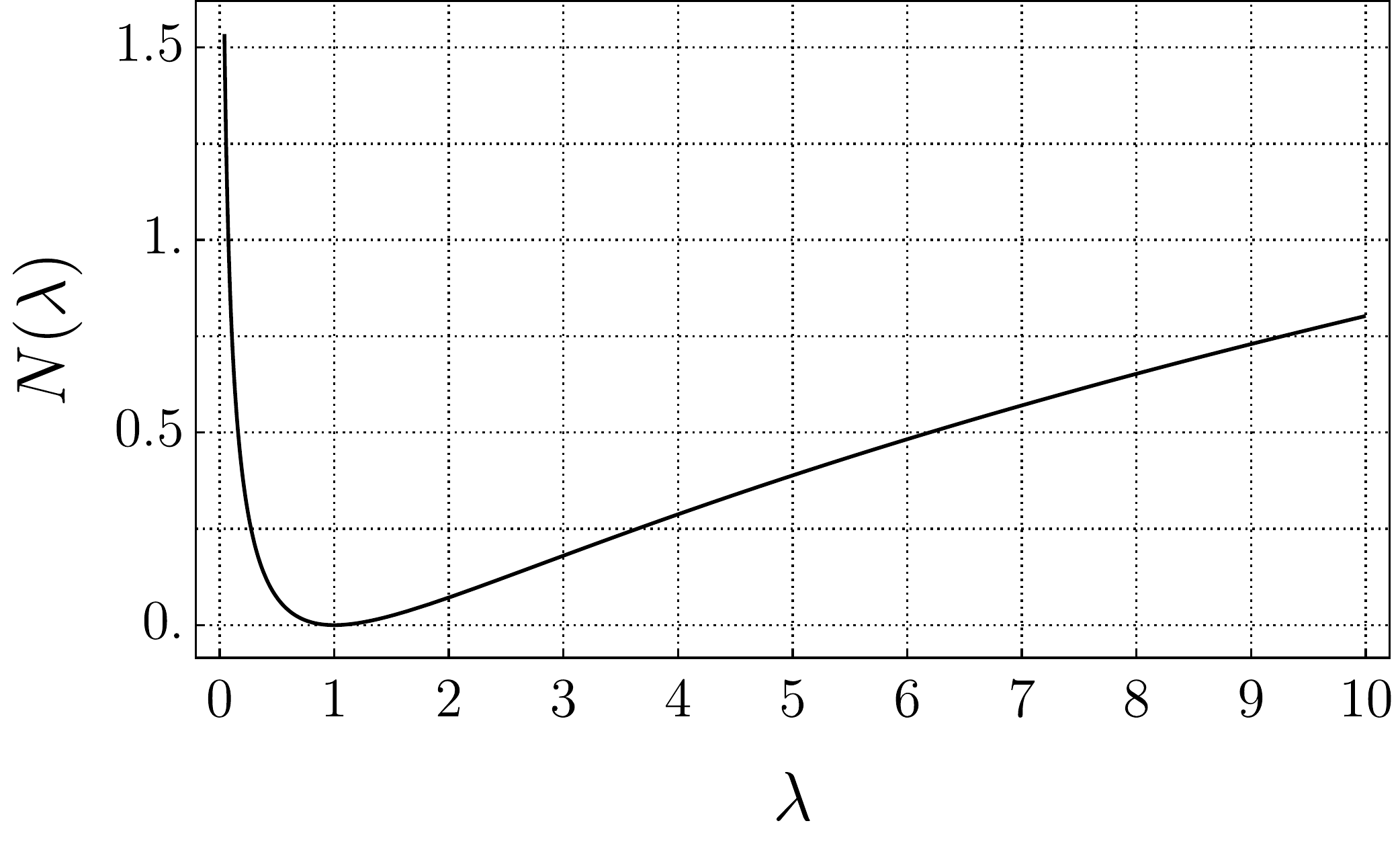}
\caption{The total number of produced particles as a function of the parameter $\lambda$ of the frame transformation.}
\label{fig1}
\end{figure}

The total number of particles produced can be found by integration of $|\gamma_{\omega\tilde{\omega}}|^2$ and it is given by
\begin{equation}\begin{aligned}
&N(\lambda)=\int_{0}^{\infty}{\int_{0}^{\infty}{|\gamma_{\omega\tilde{\omega}}|^2d\omega}d\tilde{\omega}}\\&=\frac{2 \log (\lambda ) \left(-2 \lambda ^2+\left(\lambda ^2+1\right) \log (\lambda )+2\right)+\pi ^2 (\lambda -1)^2}{\pi ^2 (\lambda +1)^2}.
\end{aligned}
\end{equation}
The plot of $N(\lambda)$ can be found in Fig.~\ref{fig1}. Notice that $N(\lambda=1)=0$ as expected (no transformation at all) and $N(\lambda)$ tends to infinity in the limits $\lambda\to 0$ and $\lambda\to\infty$.  This behaviour is expected since in these cases we are going from a non $\textrm{AdS}$ invariant frame to an invariant one.

\section*{conclusions}

We studied a dynamical process in which an observer in $\textrm{PAdS}_2$ is instantaneously transported to an isometric frame.  We showed that, if the scalar field satisfies a non-trivial boundary condition when $t<0$, then particles will be produced during the process at $t=0$ due to the break of $\textrm{AdS}$ invariance. This setup is equivalent to a sudden change of boundary conditions in $\mathbb{R}\times[0,\infty)$ Minkowski spacetime. On both cases, when the initial vacuum is not given by a field satisfying Dirichlet or Neumann boundary condition, the number of produced particles is finite.

Notice that this effect does not appear in other globally hyperbolic, maximally symmetric spacetimes, such as Minkowski and de-Sitter. A similar outcome would follow from a boost transformation in Minkowski spacetime with a point in space removed. It is worth pointing out, however, that in this case, the boost is no longer one of the symmetries of spacetime. This new result we are proposing is essentially a quantum effect since, classically, all symmetries are preserved.

\acknowledgments

We are grateful to G.~Satishchandran, R. A. Mosna and A. Saa for enlightening discussions and also thank the Enrico Fermi Institute for the kind hospitality. J.~P.~M. Pitelli thanks Funda\c c\~ao de Amparo \`a Pesquisa do Estado de S\~ao Paulo (FAPESP) (Grant No. 2018/01558-9). V.~S. Barroso thanks FAPESP (Grants No. 2018/09575-0 and No. 2016/25963-4). Finally we all thank FAPESP (Grant No. 2013/09357-9).

\end{document}